# Teaching Information Security Management Using an Incident of Intellectual Property Leakage

## Completed research paper

### Atif Ahmad
School of Computing and Information Systems
The University of Melbourne
Parkville, Victoria, Australia
Email: atif@unimelb.edu.au

### Sean B. Maynard
School of Computing and Information Systems
The University of Melbourne
Parkville, Victoria, Australia
Email: seanbm@unimelb.edu.au

### Sameen Motahhir
Center for Digital Humanities, Department of Languages and Linguistics
Information Technology University
Lahore, Pakistan
Email: sameen.motahhir@itu.edu.pk

### Moneer Alshaikh
Department of Cybersecurity
College of Computer Science and Engineering
University of Jeddah, Saudi Arabia
Email: malshaikh@uj.edu.sa

## Abstract

Case-based learning (CBL) is a powerful pedagogical method of creating dialogue between theory and practice. CBL is particularly suited to executive learning as it instigates critical discussion and draws out relevant experiences. In this paper we used a real-world case to teach Information Security Management to students in Management Information Systems. The real-world case is described in a legal indictment (T-mobile USA Inc v. Huawei Device USA Inc. and Huawei Technologies Co. LTD) alleging theft of intellectual property (trade secrets) and breaches of contract concerning confidentiality and disclosure of sensitive information. The incident concerns a mobile phone testing robot (Tappy) developed by T-mobile USA to automate testing of mobile phones prior to launch. T-mobile alleges Huawei stole the technology by copying the robot's specifications and stealing parts and software to develop its own testing robot. The incident scenario is interesting as it relates to a business asset that has both digital and physical components that has been compromised through an unconventional cyber-physical attack facilitated by insiders. The scenario sparked an interesting debate among students about the scope and definition of security incidents, the role and structure of the security unit, the utility of compliance-based approaches to security, and the inadequate use of threat intelligence in modern security strategies.

# 1  Introduction

The responsibility to protect the firm's business information resources is frequently delegated to a senior executive, typically the Chief Information Security Officer (CISO) or Chief Information Officer (CIO) (Maynard et al. 2018a). However, an informal review of information security curricula reveals, unsurprisingly, that the majority of information security teaching in tertiary education is offered from a technical rather than business perspective. Technical topics are conducive to producing IT Security Architects, IT Security Analysts, Security Technicians, Incident Responders and Forensic Investigators. Whereas programs centred on business/management aspects (i.e. Information Security Management (ISM)) are better suited to the development of CISOs or CIOs, security auditors, security policy developers, and security risk assessors (Cram and D'Arcy 2016).

Case-based Learning (CBL) instigates critical discussion, draws out relevant experiences from students, encourages questioning of accepted practices, and creates dialogue between theory and practice (Kendall and Kendall 2017). For CBL in ISM to be effective, instructors require pedagogical instruments that are developed with ISM in mind. An informal review of teaching cases shows there are very few specifically in the security domain. The few that do exist are not aimed at learning outcomes related to management practice. For example, in "Cybersecurity: the Three headed Janus", Diffee and Datta (2018) focuses on ethics, process and people in cybersecurity whereas in "Snowfall and a Stolen Laptop", McLaughlin et al. (2015) describe the incident response processes and technologies used to track a stolen laptop belonging to the Dean of the E. Philip Saunders College of Business at Rochester Institute of Technology. Therefore, we ask the following question, "How can Case-based Learning in ISM be improved?"

To answer this question we identify, apply and demonstrate the utility of our selected case as a contribution to CBL in ISM. We have identified a case that is targeted towards a critical ISM problem facing private enterprises: intellectual property (IP) theft. In the US alone, theft of trade secrets costs businesses $300 billion per year (National Bureau of Asian Research 2017). There are incentives to steal IP as competitors can gain a significant advantage on the open market by leveraging stolen IP rather than resourcing innovation themselves. Protecting IP is both complex and expensive as it involves understanding and developing mechanisms for the people, process and technology dimensions of information security, a classic problem in the discipline of Management Information Systems (Ahmad et al. 2014a).

Further, we chose the particular legal indictment because it speaks to the risks introduced by the conventional practice of delegating ISM to IT operations. Specifically, that IT operations is well suited to maintaining IT service availability but not well positioned to recognise, and therefore protect, IP. The problem is even more severe where the asset is not entirely digital or does not qualify as a conventional IT asset.

The paper is structured as follows. We present background on the teaching context before presenting details of the legal case. We then present the discussion questions, the intent of the instructors in orchestrating debate and key themes that emerged in discussions.

# 2  Background

The case is used in a subject within an on-line executive Master of Information Technology Management degree program. Students enrolling in this degree tend to be on the executive management track within their organisations. Hence, the subject is aimed at industry professionals with at least 5 years of experience in IT-related work. This section describes the design of the subject as background to better appreciate the context of the case.

The subject is designed with ISM in mind from the outset. We use the term ISM to refer to the intersection between Information Security and Management Information Systems. The following three principles inform the subject's learning objectives:
    P1. ISM aims to protect the business function of the organization and therefore must consider management practices such as strategy, risk, and policy.
    P2. ISM aims to protect sensitive data, information, and knowledge wherever it may be stored and however it may be transmitted.
    P3. ISM is essentially a busines problem with people, process, and technology dimensions.

Table 1 provides an overview of the teaching content. In the first module, we introduce ISM from an organizational perspective rather than a technology perspective to set the feel of the subject. The central theme is enterprise security, which is explained in terms of the relationship between the constructs of threat, asset, and control (see the security triplet model in Baskerville (2005)). Further, the module focuses on the primary objectives for private enterprise, that being the protection of competitive advantage (particularly IP) and IT service continuity. We show how IP theft in organisations can be facilitated through cyber means, illustrating this through legal case studies (e.g. Brooklyn Law School 2014; United States Court of Appeals for the Seventh Circuit 2019).

| Module and Content | Primary Readings |
| --- | --- |
| **1: Information Security in the Business Context** <br> The Security Triplet; Computer Security vs Information Security vs Information Assurance, The C-I-A Triangle | Ch 1, Whitman and Mattord (2017); Cherdantseva and Hilton (2015); U.S. District Court for the Western District of Washington (2014); Baskerville (2005); Ahmad et al. (2019) |
| **2: Management Practice in Information Security** <br> Overview of strategy, risk, policy, and SETA. Challenges for the CISO | Ch 2, Whitman and Mattord (2017); Maynard et al. (2018a) |
| **3: Planning & Strategy in Information Security** <br> Standards & Frameworks - ISO 27000 series, Security Strategies for Command & Control, | Ch 4, Whitman and Mattord (2017); Ahmad et al. (2014b) |
| **4: Role of Technological Controls** <br> Configuring Defences, Perimeter Defence: Firewalls / VPNs, Internal Lines of Defense: IDS, Access Control, | Ch 6 & 7, Whitman and Mattord (2017); Ahmad et al. (2014b) |
| **5: Risk in Information Security** <br> The Security Risk Management Process, Risk Identification and Assessment, Risk Control Strategies, Benchmarking and Baselining | Ch 5, Whitman and Mattord (2017); Shedden et al. (2016); Webb et al. (2014); |
| **6: Strategy in Information Security** <br> Security Triplet, Security Strategy Paradigms, Beyond the Castle Model of Defense; Governance Models | Ch 4, Whitman and Mattord (2017); Ahmad et al. (2014b); Leuprecht et al. (2016); Maynard et al. (2018b); Park et al. (2012); Tan et al. (2010); Ahmad et al. (2020) |
| **7: Policy in Information Security** <br> Policies, Standards and Practices, Policy-centric Decision Making, Enterprise vs Issue Specific Policies, Assessing Policy Quality | Ch 4, Whitman and Mattord (2017); Maynard and Ruighaver (2007); Cram et al. (2017); Maynard et al. (2011) |
| **8: SETA in Information Security** <br> SETA; Development, Delivery and Evaluation of Training | Ch 4, Whitman and Mattord (2017); Bada and Nurse (2019); Alshaikh et al. (2019); Alshaikh et al. (2018); Alshaikh et al. (2014); Shedden et al. (2011) |
| **9: Advanced Topics** <br> Knowledge Leakage; Situational Awareness in Risk Management | Ahmad et al. (2014a); Shedden et al. (2016); Thompson and Kaarst-Brown (2005); Webb et al. (2014); Ahmad et al. (2019) |

*Table 1: Overview of teaching content*

Module 2 introduces students to management security controls such as strategy, policy, risk and Security Education, Training, Awareness (SETA) and the role of the CISO (these are explored later in the subject). In module 3, students investigate how organizations plan and implement security programs using industry standards and codes of practice (e.g. the ISO2700 series ISO/IEC 2005). Module 4 discusses perimeter security technologies (e.g. firewalls, VPNs), and technologies on internal lines of defense (e.g. IDS, anti-virus, access control). The focus assisting students to understand how these technologies can be applied and combined to support a security strategy.

The security risk management process is investigated in module 5. Module 6 presents principles of security strategy as they apply to organizational infrastructure (see Ahmad et al. (2012)) and draws an analogy between the preventative mindset that lead to the development of the castle architecture and to that of modern network configurations in use today. The quality of security policy (module 7) is a research-oriented topic based on (Maynard and Ruighaver 2007). The many dimensions of quality are presented with a case study from the instructor's own experiences in a major re-drafting of a series of security policies as part of a compliance exercise. The fine distinctions between Security Education, Training and Awareness are the focus of module 8. The focus of this session is on the need to understand the root cause of poor security behaviour and to develop strategies to engage and motivate employees to align their behaviour with organizational security policy (Bada and Nurse 2019).

The final module looks at advanced topics relevant to students of management information systems. In 2020, the focus was on understanding how knowledge assets are different from conventional information and IT assets, and how said assets require innovative security strategies for protection (Ahmad et al. 2014a; Shedden et al. 2016). A second focus was on the need for organizations to develop situational awareness of the threat environment (Ahmad et al. 2019; Webb et al. 2014).

# 3 The Incident Scenario

Our incident scenario is unique as it relates to the theft of an organizational asset that has both physical and digital components. Instructors wanted students to question their definition of 'information security' and 'cybersecurity' (particularly the scope and boundaries of the definition). Further, the incident was unique as there are few if any incidents discussed in teaching cases that adopt a management information systems lens where the security incident relates to a business problem that centres on the interaction between people, process, and technology (most cases take a technology lens and focus on cases of IT service disruption, e.g. McLaughlin et al. (2015); Sipior et al. (2019); Diffee and Datta (2018)).

## 3.1 Background to T-mobile

T-mobile US Inc. is a leading mobile telephony and wireless broadband operator in the USA with annual revenues in 2019 exceeding 45 billion US dollars (Trefis Team 2020). The firm is owned by German telecommunications giant Deutsche Telekom (43%), Japanese conglomerate SoftBank Group (24%) and the public through share ownership (33%) (Shu 2020). Although T-mobile supplies technology devices like mobile phone handsets to its customers, it does not manufacture these devices but rather uses a competitive tender process to procure these from suppliers. Despite the competitive tender process, sub-standard devices were resulting in a high rate of return from its customer base. Poor device reliability was a significant competitive disadvantage for T-mobile (Salman 2012).

## 3.2 The Competitive Advantage

In 2007, T-mobile engineered a robot arm nicknamed 'Tappy' to test software glitches in the mobile phones and tablets procured from suppliers (Cheng and Keane 2019). The robot arm features a rubberized tip used to operate device screens (e.g. press buttons, push rollerballs, navigate touch screens) and is driven by software that tests a wide range of scenarios (typing, playing music, making calls, gaming, web browsing, downloading applications) to test the device's functionality for glitches (e.g. memory leaks, stalls, freezing, responsiveness). A video camera positioned strategically records the device's responses to the robot arm's commands to capture said glitches. After each test, Tappy produces a detailed report showing how the device performed and where the glitches occurred. Only devices that perform perfectly for 24 consecutive hours are released to the public.

According to T-mobile, since the introduction of Tappy, device returns were reduced by up to 75% (Salman 2012). As a result of this significant competitive advantage (no other firm had a testing robot at this point), T-mobile competitors began developing their own testing robots. Huawei China's efforts to build 'xDeviceRobot' did not prove fruitful (Wamsley 2019). In fact, Huawei devices had the poorest record among all the manufacturers supplying T-mobile. Huawei asked T-mobile to license the technology to them, however T-mobile refused the request. It was at that point that Huawei China enlisted the help of Huawei USA to collect intelligence about Tappy. The request for intelligence eventually snowballed into an incident of industrial espionage, and a series of legal indictments.

## 3.3 The Incident of Theft

T-mobile had designed strict security protocols to protect Tappy (U.S. District Court for the Western District of Washington 2014). Tappy was housed under tight security in a dedicated lab at the firm's headquarters in the greater Seattle area. Initially, only T-mobile employees could operate Tappy. However, later, T-mobile relaxed the restrictions to allow employees from its supplier partners to operate the robot arm as well. Prior to accessing Tappy, T-mobile required all suppliers including Huawei to sign contracts prohibiting them from taking any photos, videos, or using any knowledge gleaned about Tappy during the testing process to reverse engineer Tappy.

The legal indictment describes the incident in detail as a sequence of four key events (U.S. District Court for the Western District of Washington 2014):

*Event A: The Line of Questioning: "On several occasions in 2012 and continuing to 2013, Huawei employees asked T-Mobile personnel detailed questions about the testing robot. These questions frequently concerned the conductive tip on the end of the 'end effector' – a metal plate that affixes to*

*the bottom of the robot arm. Indeed, as time went on Huawei's questions became more pointed and intrusive about the exact operational details of the robot. T-Mobile personnel did not provide answers to these questions about T-Mobile's proprietary technology."* [page 10 of 22, section 43].

*Event B: The Photographs:* *"….one day after Mr. Wang had just been told he could not enter the lab, Mr. Xiong and Ms. Lijingru secretly escorted Mr. Wang into the T- Mobile testing lab to photograph the testing robot. Using a smartphone, Mr. Wang took at least seven photographs of the testing robot, in violation of the Clean Room Letter's explicit prohibition on photography in the lab… That night, Mr. Wang forwarded the photographs he had taken of the robot to Huawei's R&D team in China."* [page 11 of 22, section 48, 50].

*Event C: Theft of a Physical Artifact:* *"While alone in the lab, Mr. Xiong attempted to hide one of these end effectors out of the view of the security camera behind a computer monitor. Three hours later, while glancing repeatedly at the security camera, Mr. Xiong moved the end effector from behind the monitor and slipped it into his laptop bag. Mr. Xiong then carried the laptop bag out of the testing chamber… Mr. Xiong took the end effector to Huawei Device USA's offices and used it to provide measurements to Huawei China's R&D division during a conference call."* [page 11 of 22, section 55, 57].

*Event D: Theft of Software:* *"Sequence files control the movements of the proprietary testing robot. These files, which are themselves proprietary trade secrets, are present on the computers in the testing lab. On information and belief, Huawei employees accessed and sent proprietary sequencing files via email to others at Huawei. They were explicitly forbidden from doing so by T-Mobile."* [page 13 of 22, section 61,62].

# 4 Student Engagement with the Incident Scenario

## 4.1 Context of the Incident Scenario

The T-mobile v Huawei incident is presented in the very first of the teaching modules. Students are given background on the very real and pervasive problem of IP theft after which the Huawei vs T-mobile case is presented with a short 1-minute video clip from T-mobile showing employees describing the bugs they catch using Tappy (T-Mobile 2013) and the following preamble:

*"USA-based T-Mobile initiated legal proceedings against Chinese mobile manufacturer Huawei alleging the latter stole the designs and parts to its proprietary phone testing robot technology (known as 'Tappy'). Tappy was an internally developed and refined piece of equipment that developed and refined piece of quality assurance equipment that was instrumental in removing defects from mobile handsets.*

*Tappy was placed in a secure location and carefully guarded with strict physical and legal security protocols to 'protect the trade secrets invested in the robot, its component parts, specifications, software and functionality'. The firm alleges that Huawei stole the technology by copying the robot's specifications and stealing parts, software and trade secrets to develop its own testing robot."*

## 4.2 Discussion Questions, Instructor Intent and Student Response

Students are presented with a link to the legal indictment (U.S. District Court for the Western District of Washington 2014) and four quotes from the legal indictment (events A to D as above in section 3.3). Students are then taken to an online discussion forum where they are presented with three questions. We discus each question, instructor intent and student responses below:

**Question 1:** Would you call this an information security incident? Why or why not?

The scenario is interesting to the instructors because Tappy was not a conventional IT asset and the incident was not the result of a conventional cyber-attack. The first question encourages students to question the utility of the technology perspective in identifying risks and estimating busines impact. The intent is for students to adopt the business perspective once they come to recognize the limitations of the technology perspective.

In their response to the first question, students focused on the boundary between accidental leakage vs industrial espionage pointing out that T-mobile took active measures to protect Tappy and Huawei purposively and maliciously breached security protocols to steal IP. Some students focused on implicit evidence of Huawei's planning and preparation as an argument for declaring the events an incident. Others debated whether it was an information security incident or just a security incident since Tappy was not a digital asset and had a distinct physical component. For example, one student stated:

*"By my understanding of the definition – yes, this was indeed an "information security incident" as data, IP, 'information' etc. was stolen; however, physical items were also stolen leading me to also classify the incident more broadly as a 'security incident'."*

**Question 2:** Which organisational function or department would be responsible for preventing such an incident from occurring?

The second question encourages students to further explore the implications of adopting a business perspective as opposed to a technology perspective to security incidents. Specifically, the fact that in most organizations it would be unclear which unit or department would be responsible for the security of Tappy given it is not a laptop, server, or other conventional IT device (i.e. the security unit in many organizations is mainly responsible for IT security only). Many organizations do not retain the equivalent security unit in the physical and human resource domains. Physical security usually consists of outsourced building managers and security guards that manage doors, access control, etc. They will likely consider monitoring Tappy outside of their responsibility. The main responsibilities of Human Resources are recruitment, selection, performance management and onboarding and offboarding employees.

In their response to the question, most students suggested that the responsibility for preventing such an incident was shared by all employees and that the senior management were ultimately accountable for the failure of security. For example:

*"Simply put, information security is everyone's responsibility, in this case different functions from left to right (Info sec, HR, Risk, IT, Physical Security, Sourcing & Procurement etc) and top (management) to bottom (individual T-mobile staff) do play a role to safeguard the asset."*

*"I do believe the ultimate responsibility falls on senior management. They should have had an effective information assurance strategy in place. They should have commissioned a risk assessment; this would have identified which organisational functions and departments were responsible for which assets and in which capacity. They should also have had processes, policies, training, controls etc. in place to protect valuable assets like Tappy."*

At this point the instructors directed students to consider certain aspects of the scenario. Using an organizational chart, the instructors explained that the structural divisions between IT, legal and HR make it difficult to translate a shared responsibility into a focused response to a security incident. Further, that in practice securing IP is not a priority for legal, HR and physical security units. The instructors' comments lead to a discussion on the empowerment of the CISO as well as queries on the possibility of having the legal, HR and physical security units report to a broader enterprise-wide Chief Security Officer to overcome structural divisions. For example:

*"Are these CSO peering with their CIO and CTO counterparts? In my experience, even IT struggles to get a real position of power at a board level (often the CIO sits under a COO who is a board member) diluting the influence within the business. Considering security is closely intertwined with risk management, have you seen examples of security and risk being coupled and managed under the same management structure with board level influence?"*

Another key thread of discussion debated the utility of compliance in security. Students reflected on their experience working in regulated environments:

*In highly regulated industries such as financial services, it is my experience that there is compliance fatigue that can undermine security initiatives bottom-up. Could this be solved through more effective communication of the underlying security imperative? This could stop the sometime held view of compliance being a barrier to someone doing their job and little more. This for me then ties back to either the weight of the CISO role in the organisation and the relationships held with the C-Suite executives and the Board...*

And:

*I feel for people working in the regulated industries as I am having similar experience. What I've found after all is whether an organisation has good security people (both leaders and operations on the ground) to pivot the control environment from compliance driven to security driven/embedded. Compliance itself actually provides a good foundation... Through the compliance process e.g. control compliance monitoring, the controls (formal, technical) could be more polished and agile to be effective to fit for purposes. Compliance is not everything and it will never be good enough, but it can form a baseline of a healthy cyber ecosystem.*

**Question 3:** How can organisations like T-Mobile prevent such incidents from taking place? What countermeasures would you put in place?

The intent of question 3 was to shift the discussion from that of organizational response to that of organizational prevention of a security incident. Given the scenario describes a security incident that doesn't fit the definition of a conventional scenario of cyber-attack, students would not be able to use conventional technology controls and responses (e.g. encrypt files, compartmentalize digital environments firewalls, detect malware using intrusion detection systems, anti-virus software) and would be lead to question the underlying organizational strategy for handling incidents.

At first students suggested the conventional improvements in security measures (better monitoring, better training, tighter physical access to labs). Some students simply argued at a high-level, suggesting better information assurance strategies, auditing of management executives etc.

However, some of the more experienced executive students pointed out that the attack was facilitated by insiders that were authorized to be in the general vicinity of Tappy. Further, that since the adversary had passed through the 'shield', then the remaining operational security measures can be more easily circumvented. For example:

*I don't think a single specific organisational function or department is or could be responsible for preventing the incident from occurring - Huawei was invited into the 'Tappy' lab by T-Mobile to collaborate on a product that would mutually benefit both parties. Standard security functions (physical, contractual etc.) in place were relatively non-specific to their partnership and I imagine the T-Mobile employees themselves were required to adhere to similar security protocols.*

The conversation took an interesting turn when the query was raised as to why T-mobile had not identified Huawei as a threat given its prior track record in IP theft and its declared interest in acquiring a license to access Tappy:

*If T-Mobile recognised Tappy as a competitive advantage with significant IP, they should have limited interaction to Tappy with all non-employees where they thought the risks (losing the competitive advantage obtained through IP in Tappy) exceeded the benefit (faster approved phones to consumers and/or better quality phones than other carriers). To me, this seems like a strategic business decision that could then ultimately define the other security measures in place. If the cost to the business was too great, interactions should be minimised, and protection of the IP should be a primary goal that trumps quicker debugging.*

*I completely agree that increased monitoring and escalation of suspicious activity would have resulted in a better outcome, and ideally these kinds of procedures could have been established before inviting potential threats to dabble with T-Mobile's intellectual property. If T-Mobile had policies and procedures in place to mandate nondisclosure and confidentiality agreements, is it too much to ask that they also analyse threats and make informed business decisions before sending these prior to future engagements?*

The instructors weighed in again on the conversation. This time they provided students with a means of prioritizing threat actors to inform preventative strategies:

*… part of conducting a threat assessment on a potential competitor is to ask two basic questions. Does the competitor have a reason to steal IP? Does the competitor have the means or capability to steal the IP? These two factors (means and motivation) narrow down the pool of potential threat actors to the critical ones. Had T-mobile done that, they almost certainly have placed Huawei at the top of their list. Wouldn't a natural outcome of this observation be to actively monitor all interactions with Huawei and take swift action if there was any indication of a breach in progress?*

Students were quick to internalize this logic and apply it to develop new insights:

*I didn't realise there was quite a bit of public discussion around Huawei's ties to the Chinese state and potential espionage back in 2011 which would have been uncovered in the threat assessment, as you mentioned. So, using means and motivation a T-Mobile threat assessment might play out a little like this in 2012:*

    **Apple:**    <u>Means</u> – *High, huge market cap and cash pile.*
                         <u>Motivation</u> – *Low, they've probably got their own equivalent device.*

    **HTC:**      <u>Means</u> – *Medium, average manufacturer.*
                             <u>Motivation</u> – *Medium, focused on cheaper devices, user experience less relevant.*

***Samsung:*** <u>Means</u> – High, chaebol, money and politically connected in South Korea.
<u>Motivation</u> – Very High, they want any advantage to improve their high-end market and have open cases of design theft.

***Huawei:*** <u>Means</u> – Very High, connected to Chinese state, maybe involved in espionage.
<u>Motivation</u> – High, they want any advantage to improve their high-end market.

*Using the findings around potential espionage, I'd agree Huawei would be on the top of the list and I'd want to actively monitor any suspicious activity. I'm guessing, based on the above assumptions, I'd also want to pay closer attention to suspicious activity from Samsung over HTC or Apple?"*

## 5 Discussion

The underlying objective of our teaching is to develop a new generation of Information Systems practitioners in ISM that will have the foundational knowledge to protect business information resources (see the first principle that informs our learning objectives). The first step in developing this knowledge is for students to adopt the business or organisational perspective and develop an understanding of its implications. In order to achieve this goal, our most significant challenge was to encourage students to question the limitations of the existing technology perspective that is deeply entrenched in the prevailing discourse across media, in the majority of the academic literature (ISM research is conducted largely by technologists as opposed to business information systems researchers), and even in industry itself (most industry practitioners in ISM have technology backgrounds).

As Ahmad et al. (2014a) points out, an interesting dilemma emerges from the delegation of ISM to technology operations. That being, that IT personnel adopt a technology lens rather than a business lens to their routine operations. In most cases, business objectives and technology objectives align well, for example in the case of IT service availability. However, in some cases the objectives do not align. This is particularly where the asset is competitively sensitive knowledge or information. Here, the sensitivity is known to the business and IT operations are not well positioned to recognise, and therefore protect, IP. The problem is even more severe where the asset is not entirely digital or does not qualify as a conventional IT asset. Tappy is precisely this kind of asset.

Executive students learn best when they leverage their own experiences and engage in problem-based learning, action-learning and reflective learning in a community of practice (Wuestewald 2016). In our experience with this scenario, we found executive students responded very well because it presented them with an opportunity to reflect on their practical experience. Overall, fifteen of the twenty-two students participated in the conversation and the thread had 52 posts which is more than twice the number of posts in any discussion thread in the subject. Further, as exemplified in the response to the final question, more advanced students demonstrated an ability to apply the business perspective to develop new practical insights, which suggests the case fulfilled its objectives.

## 6 Summary and Future Work

The T-mobile v Huawei case provides a useful basis for teaching Information Systems students about ISM. The case is a useful addition to the limited number of cases in ISM as it relates to a business asset that has both digital and physical components that has been compromised through an unconventional cyber-physical attack facilitated by insiders. The case sparked an interesting debate among executive students about the scope and definition of security incidents, the role and structure of the security unit, the utility of compliance-based approaches to security, and the inadequate use of threat intelligence in modern security strategies. In the second phase of this project we will be conducting a second and much larger study of executive students to confirm the utility of the case in teaching ISM.

## Copyright